\documentclass[11pt]{article}
\usepackage{geometry}
\usepackage{graphicx}
\usepackage{amssymb}
\usepackage{amsmath}
\usepackage{epstopdf}
\newcommand{\yd}{\frac{1}{2}}
\newcommand{\yc}{\frac{1}{4}}

\newcommand{\sla}{\sqrt{\la}}
\newcommand{\hb}{\hbar}

\newcommand{\lef}{\langle}
\newcommand{\ra}{\rangle}

\newcommand{\mo}{m\omega}

\newcommand{\al}{\alpha}
\newcommand{\la}{\lambda}
\newcommand{\ep}{\epsilon}
\newcommand{\si}{\sigma}

\newcommand{\ga}{\gamma}

\newcommand{\p}{\partial}

\newcommand{\be}{\begin{equation}}
\newcommand{\ee}{\end{equation}}
\newcommand{\bi}{\begin{itemize}}
\newcommand{\ei}{\end{itemize}}
\newcommand{\bn}{\begin{enumerate}}
\newcommand{\en}{\end{enumerate}}
\newcommand{\ben}{\begin{eqnarray}}
\newcommand{\een}{\end{eqnarray}}

\title{Complexifier  Versus Factorization and Deformation Methods For Generation of  Coherent States of  a 1D NLHO \\ I. Mathematical Construction}
\author{R. Roknizadeh$^{1,2}$, H. Heydari$^2$\\
\small{ 1. Department of Physics, Quantum Optics Group, University of Isfahan, Isfahan, Iran}\\
\small{2. Physics Department, Stockholm University 10691 Stockholm Sweden}}
\begin{document}
\maketitle
\begin{abstract}
Three methods: complexifier, factorization and deformation, for construction  of coherent states are presented for one dimensional nonlinear harmonic oscillator (1D NLHO). Since by exploring the Jacobi polynomials $P_n^{a,b}$'s,  bridging  the difference between them is possible, we give here also the exact solution of Schr\"odinger equation of 1D NLHO in terms of Jacobi polynomials.
 \end{abstract}
\section{Introduction}

For description of the dynamics of a  non-linear harmonic oscillator (NLHO) \cite{Laksh}, we will encounter with three  types of annihilation operators, their eigenfunctions can be considered as coherent states: (i) The annihilation operator, which is obtained from {\it complexifier}, introduced by Thiemann \cite{Thiem}; (ii) The annihilation operator, which is constructed out of the  annihilation operator of simple harmonic oscillator (SHO), by using the method known in the nonlinear coherent states theory \cite{Manko,Rokni1}, and (iii) The annihilation operator which  appears in the factorization of the shape invariant potentials \cite{Carin 1}.

  Our work is divided in two parts:  The present contribution dealing with the  mathematical construction of three types of  coherent states of a nonlinear harmonic oscillator.

 In the second part, the relation between them, their properties, and some physical realization, will be given by exploring the properties of Jacobi polynomials (instead of Hermitian polynomials in the SHO theory).

The paper is organized as follows: In section \ref{2}, by introducing a canonical transformation, we give a prescription for quantization of classical Hamiltonian, by which the Schr\"odinger equation takes a familiar form of a particle in a $\tanh^2$-potential. Section \ref{3} is devoted to solving the Schr\"odinger equation, and finding  the eigenvalues  and  explicit form of eigenfunctions of the Hamiltonian. In section  \ref{4}, classical and quantum complexifier for this system are derived. Finally in section \ref{5}, three types of coherent states related to the  different annihilation operators will be presented.


\section{Schr\"odinger equation of 1D NLHO }\label{2}

The equation of one dimensional nonlinear harmonic oscillator is given by \cite{Laksh},
\begin{equation}\label{ne}
m[(1+\la x^2)\ddot{x}-\la x\dot{x}^2+\omega^2 x]=0,
\end{equation}
where $\omega$ is the angular frequency, $\la\ge0$ is a real parameter with physical dimension $(length)^{-2}$ and for $\la=0$ the equation of motion of an oscillating  particle with mass $m$  is obtained.
The equation of the 1D NLHO can be obtained from the following Lagrangian \cite{Carin 1}
\begin{equation}
L=\yd\left(\frac{m}{1+\la x^2}\right)(\dot{x}^2-\omega^2 x^2).
\end{equation}
The canonical momentum is given by
\begin{equation}
p=\frac{\p L}{\p \dot{x}}=\frac{m\dot{x}}{1+\la x^2}
\end{equation}
and from this equation we get
\begin{equation}
\dot{x}=(1+\la x^2)p/m, \quad \dot{p}=-\frac{\la xp^2}{m}-\frac{m\omega^2x}{(1+\la x^2)^2}.
\end{equation}
Moreover, the Hamiltonian can be obtained  as follows
\begin{equation}\label{Hamil}
H=p\dot{x}-L=\frac{(1+\la x^2)p^2}{2m}+\frac{m\omega^2x^2}{2(1+\la x^2)}
\end{equation}
Note also that $\dot{p}=-\p H/\p x$.
From the energy conservation $H=E$, $H$ describes an oscillator with an  amplitude dependent frequency,
\be
x(t)=A\sin(\Omega t+\Phi), \quad A=\sqrt{\frac{2E}{m\omega^2-2E\la }},\quad
 \Omega=\sqrt{\frac{2E}{mA^2}}=\sqrt{\frac{m\omega^2-2E\la}{m}}.
\ee
Thus, by  using equation (\ref{ne}), we get
\be
\frac{\Omega}{\omega}=\frac{1}{\sqrt{1+\la A^2}},\quad E=\frac{m\omega^2}{2\la}\left[1-\frac{1}{1+\la A^2}\right].
\ee
Now, by define the following canonical transformation
\begin{equation}\label{trans}
\sqrt{1+\la x^2}p=P,\quad \frac{1}{\sqrt{\la}}\sinh^{-1}(\sqrt{\la}x)=X
\end{equation}
 we get
\be \label{P}
\dot{X}=\frac{P}{m},\quad \dot{P}=-\frac{m\omega^2}{\la}\frac{\sinh\sla X}{\cosh^3\sla X}=-\frac{\p}{\p X}\frac{m\omega^2}{2\la}\tanh^2(\sqrt{\la}X).
\ee
And for these  transformations, it is easy to check the following Poisson brackets:
\begin{equation}
\{x,p\}_{(X,P)}=1=\{X,P\}_{(x,p)},\quad \lim_{\la\to 0}X=x,\;\;  \lim_{\la\to 0}P=p,
\end{equation}
where
\begin{equation}\label{pois}
\{f,g\}_{(x,p)}=\left(\frac{\p f}{\p x}\frac{\p g}{\p p}-\frac{\p f}{\p p}\frac{\p g}{\p x}\right),\quad\text{and}~~~\frac{\p X}{\p x}=(1+\la x^2)^{-1/2}.
\end{equation}
The Hamiltonian in the  new coordinates is given by
\begin{equation}\label{TH}
H(X,P)=\frac{P^2}{2m}+\frac{m\omega^2}{2\la}\tanh^2(\sqrt{\la}X).
\end{equation}
The  form of potential energy in this coordinate system  has  clear relation to the force equation in (\ref{P}),
 and also $\dot{P}=-\p H/\p X$.

 {\bf Remark.} Even though  $H$ looks like  the Hamiltonian of a particle in a potential given by a $\tanh^2$ function (or
$\tan^2$-function if $\la<0$), but   under the canonical transformation
$x\to X$ and $p\to P$, the relation between the  kinetic  and potential energy is more subtle and the former is  not given simply by $P^2/2m$.

Before quantization of the Hamiltonian and writing  the Schr\"odinger equation, we have to consider the following  relations in Schr\"odinger representation: $x\to\hat{x}$,
$p\to\hat{p}={-i\hb d/dx}$ and $[\hat{x},\hat{p}]=i\hb$. Thus there are many  representations for $\hat{P}$ regarding the Taylor expansion of $\sqrt{1+\la x^2}$. Two of these representations are given by,
‚Äé‚Äé\be\label{ke}
\hat{P}_1=-i\hb\sqrt{1+\la x^2}\frac{d}{dx},\quad\text{with}~~~ \hat{P}_1^2=-\hb^2\left(\la x\frac{d}{dx}+(1+\la x^2)\frac{d^2}{dx^2}\right),
\ee
and
‚Äé‚Äé\be
\hat{P}_2= -i\hb\frac{d}{dx}\sqrt{1+\la x^2},\quad \text{with}~~~\hat{P}_2^2=-\hb^2\left(\la+3\la x\frac{d}{dx}+(1+\la x^2)\frac{d^2}{dx^2}\right).
\ee
However the commutation relations of these two representations with $X$ has the same result,
\[[\hat{X},\hat{P_1}]=\left[\frac{1}{\sqrt{\la}}\sinh^{-1}(\sqrt{\la}x),-i\hb\sqrt{1+\la x^2}\frac{d}{dx}\right]=i\hb,\]
\[ [\hat{X},\hat{P_2}]= \left[\frac{1}{\sqrt{\la}}\sinh^{-1}(\sqrt{\la}x),-i\hb\frac{d}{dx}\sqrt{1+\la x^2}\right]=i\hb.\]
Thus both  representations can be realized as $-i\hb (d/dX)$.
 But  we should choose  one of them from the outset, because these two representations do not commute and we would have the ordering problem of operators as is usual in quantization theories. On the other hand  the  similarity transformation
\be
e^{f({x})}\hat{P}_1e^{-f(x)}=\hat{P}_2,\quad f({x})=\ln(1+\la x^2)^{-1/2}
\ee
 relates $\hat{P}_1$ and $\hat{P}_2$. And  the Hamiltonians
$
 \hat{H}_j=\frac{\hat{P}_j^2}{2m}+V(X)$ with $j=1,2
$
 are also related by the same similarity transformation,
 ‚Äé‚Äé\be
 e^{f({x})}\hat{H}_1e^{-f(x)}=\hat{H}_2.
\ee
Their eigenvalues are identical  and their eigenfunctions are related by $e^{f({x})}\psi_1=\psi_2$.
By using the  first representation  $\hat{P}_1\equiv\hat{P}=-i\hb(d/dX)$, we get the Schr\"odinger equation
\begin{equation}\label{SE}
\hat{H}\varphi(X)=[-\frac{\hb^2}{2m}\frac{d^2}{dX^2}+\frac{m\omega^2}{2\la}\tanh^2(\sqrt{\la}X)]\varphi(X)=
E\varphi(X).
\end{equation}

\section{Solution of the Schr\"odinger equation of NLHO}\label{3}

The Schr\"odinger equation for $\tanh^2$-potential (\ref{SE}) can be solved analytically by using the methods developed in \cite{Mors, Pertsch}, and in this section we review the  the solution for later references.

\subsection{Hypergeometric differential equation}

First by defining three  dimensionless parameters
\begin{equation}\label{def}
\sqrt{\la}X=z, \quad \frac{m^2\omega^2}{\hb^2\la^2}=v, \quad \frac{2mE}{\la\hb^2}=\epsilon.
\end{equation}
Then the equation (\ref{SE}) can be rewritten as
\begin{equation}
\frac{d^2\varphi(z)}{dz^2}-[v\tanh^2z-\epsilon]\varphi(z)=0.
\end{equation}
Now, if we consider the following transformation
\begin{equation}
z=\tanh^{-1}\frac{i}{2}\frac{1-2u}{\sqrt{u(1-u)}}=\sinh^{-1}i(1-2u),
\end{equation}
then we obtain  $u=(1/2)(1+i\sinh z)$, and $d^2\varphi/dz^2=-u(1-u)(d^2\varphi/du^2)-(1/2)(1-2u)(d\varphi/du)$.
Therefore we have the following second order differential equation,
\begin{equation}
u^2(1-u)^2\frac{d^2\varphi}{dz^2}+\yd u(1-u)(1-2u)\frac{d\varphi}{du}-\left[\frac{v}{4}(1-2u)^2+\epsilon u(1-u)\right]\varphi(u)=0.
\end{equation}
Next, we define a new function
\be\label{Phi}
\Phi=u^{1/4}(1-u)^{1/4}\varphi,
\ee
 the first order differentiation will be removed from the equation and we get
\begin{equation}
u^2(1-u)^2\frac{d^2\Phi}{dz^2}+\left[(\ep+\yc-v)u^2-(\ep+\yc-v)u+\frac{3}{16}-
\frac{v}{4}\right]\Phi(u)=0
\end{equation}
which could be  transformed  to a  hypergeometric equation by defining the following function
\begin{equation}\label{y}
y(u)=u^{-\si}(1-u)^{-\si}\Phi,
\end{equation}
where $ \si=\yd+\yd\sqrt{\yc+v}$ and $\si^2-\si=(-3/16)+(v/4)$. After some calculus, we get
\begin{equation}\label{hg}
u(1-u)\frac{d^2 y}{du^2}+2\si(1-2u)\frac{dy}{du}-[2\si^2-\epsilon-\frac{1}{8}-\frac{v}{2}]y(u)=0.
\end{equation}
This has the  desired form of a hypergeometric differential equation with well known  solutions.
\subsection{The general solutions and energy eigenvalues}

The equation (\ref{hg}) has the  general form
\be\label{hge}
u(1-u)\frac{d^2w}{du^2}+[\gamma-(\al+\beta+1)u]\frac{dw}{du}-\la\beta w=0
\ee
which is satisfied by Gauss hypergeometric functions for $|u|<1$. This equation has exactly three singular points: $u=0, u=1$, and $u=\infty$. When none of $\gamma, \gamma-\al-\beta$,  or $\al-\beta$ is equal to an integer, two fundamental
solutions are known for $u$ near each of the three singular points of the differential equation.  Because of the
 Schr\"odinger boundary condition, we study the solutions near $u=\pm\infty$.
Two linearly independent solutions for $|u|\to\infty$ is given by \cite{Abram},
\begin{equation}\label{sol1}
u^{-\al}\,_2F_1(\al,1+\al-\gamma; 1+\al-\beta;u^{-1}),\quad
u^{-\beta}\,_2F_1(\beta,1+\beta-\gamma; 1+\beta-\al;u^{-1}).
\end{equation}
There is an equivalent  representation for each of the above solutions, which we give here  for later applications
\be\label{sol2}
u^{\beta-\gamma}(u-1)^{\gamma-\al-\beta} {_2F_1}(1-\beta,\gamma-\beta; \al-\beta+1;u^{-1}),~~
u^{\al-\gamma}(u-1)^{\gamma-\al-\beta} {_2F_1}(1-\al,\gamma-\al; \beta-\al+1;u^{-1}).
\ee
The general form of $_2F_1$ has the following expansion
\be\label{F}
_2F_1(a,b,c;u)=\sum_{n=0}^{\infty}\frac{(a)_n(b)_n}{(c)_n}\frac{u^n}{n!},
\ee
 where e.g.,
$(a)_n=a(a+1)(a+2)\cdots(a+n-1)=[(a+n-1)!]/[(a-1)!],\; (a)_0=1$.
The Gauss serie reduces to a polynomial of degree $n$ in $u$, when $\al$ or $\beta$ is equal to $-n$,
$n\in\mathbb{N}_0$.\\
The  differential equation (\ref{hg}) is like (\ref{hge}) with  the following parameters:
\begin{equation}\label{parameters}
\gamma=2\si,\quad \al+\beta+1=4\si,\quad \al\beta=2\si^2+\epsilon-\frac{1}{8}-\frac{v}{2}.
\end{equation}
From the last two relations one can easily find
\begin{equation}\label{ab}
\al=2\si-\yd-\sqrt{-\epsilon+v},\quad \beta=2\si-\yd+\sqrt{-\epsilon+v}.
\end{equation}
Because of the second relation in (\ref{parameters}), the sign of $\sqrt{v-\epsilon}$ in expressions of $\al$ and $\beta$ should be opposite, and  it seems that they could be chosen conversely. But since some restriction will be inserted  on 
$\beta$,   we have to choose the sign as given above.\\
According to the relations (\ref{Phi}), (\ref{y}),
\begin{equation}\label{ux}
\varphi(z)=u^{\si-1/4}(1-u)^{\si-1/4)}y(u),\end{equation}
which is defined as before by
\be\label{xu}
u=\yd(1+i\sinh z)\equiv\yd(1+iq),\quad u(1-u)=|u|^2=(1+q^2)/4=[\cosh^2z]/4
\ee
so that
\begin{eqnarray}\label{seri}
\varphi(z) =u^{\si-1/4}(1-u)^{\si-1/4)}\Big[c_1u^{-\al}\,_2F_1\left(\al,1+\al-\gamma; 1+\al-\beta;u^{-1}\right)\nonumber\\
+ c_2 u^{-\beta}\,_2F_1\left(\beta,1+\beta-\gamma; 1+\beta-\al;u^{-1}\right)\Big].
\een
If $z\to\pm\infty$, then $q\to \infty$, $u^{-1}\to 0$,  and
  we have $_2F_1(u^{-1}=0)=1$, thus
\begin{equation}
\lim_{z\to\pm\infty}\varphi(z)=u^{\si-1/4}(1-u)^{\si-1/4)}
\Big[c_1u^{-\al}+c_2u^{-\beta}\Big].
\end{equation}
If  we absorb temporary  the phase factor of  $u^{-\al}$, $u^{-\beta}$ in $c_1, c_2$ and set $|u|$ instead of $u$, then the boundary condition will be satisfied for $u=-\infty$ and $u=+\infty$ simultaneously. Now, by using the definition of $\si$ in (\ref{y}) and its relation with $\al,\beta$  in (\ref{ab})
we get
\begin{equation}
\lim_{z\to\pm\infty}\varphi(z)=c_1\yd[\sqrt{1+q^2}]^{-(\sqrt{v-\ep})}
+c_2\yd[\sqrt{1+q^2}]^{(\sqrt{v-\ep})}.
\end{equation}
The second term goes  to infinity as $z\to\pm\infty$, so we have to set $c_2=0$.
The serie in (\ref{seri}) converges  if $|u|^{-1}<1$, or equivalently,  if $(1+q^2)/4>1$, then $|q|>\sqrt{3}$. The requirement for the wave function being  everywhere finite, forces us  to the analytical continuation of
 $\varphi(u)$ in the interval $|u|<1$ or $q$ in $(-\sqrt{3},\sqrt{3})$, (the endpoints are not included).  The idea is to define a  transformation  to map as large a region of the complex plane as possible onto
the interval  $ \yd(1- i\sqrt{3})<u<\yd(1+i\sqrt{3})$ or $(e^{-i\pi/3}<u<e^{i\pi/3})$.  If $\arg(-w)<\pi$, then the analytic continuation of  $ _2F_1(w) $  is given by \cite{nist}
\ben\label{ac}
_2F_1(a,b;c;w)&=&\frac{\pi}{\sin(\pi(b-a))}\Big[\frac{\Gamma(c)(-w)^{-a}}{\Gamma(b)‚Äé\Gamma‚Äé(c-a)} {_2F_1}(a,a-c+1;a-b+1;\frac{1}{w})\nonumber\\
&-&\frac{\Gamma(c)(-w)^{-b}}{\Gamma(a)‚Äé\Gamma‚Äé(c-b)}  {_2F_1}(b,b-c+1; b-a+1;\frac{1}{w})\Big].
\een
where
\[\frac{\pi}{\sin(\pi \eta)}=\Gamma(\eta)\Gamma(1-\eta).\]
After setting $c_2=0$ in (\ref{seri}) and $a=\al$, $b=\al-\gamma+1$, $c=1+\al-\beta$ and $w=u^{-1}$ in (\ref{ac}), we get
\be\label{seri2}
\varphi(z)=(-1)^{\al}c_1u^{\si-1/4}(1-u)^{\si-1/4)}\Big[ A\;{_2F_1}(\al,\beta;\gamma;u)
-Bu^{1-\gamma}\;{_2F_1}(\al-\gamma+1,\beta-\gamma+1; 2-\gamma;u)\Big],
\ee
where
\be\label{AB}
A=\frac{\pi}{\sin 2\pi\si}\frac{\Gamma(1+\al-\beta)}
{\Gamma(\al-\gamma+1)‚Äé\Gamma‚Äé(1-\beta)},
\ee
\be
B=\frac{\pi}{\sin 2\pi\si}\frac{\Gamma(1+\al-\beta)}{\Gamma(\al)‚Äé\Gamma‚Äé(\gamma-\beta)}.
\ee
We have to handle two problems in the above transformation: First, due to  the presence of
$\Gamma(1+\al-\beta)$ in the numerators
 and $\sin 2\pi\si$  in denominators of  $A$ and $B$,  the cases $\gamma=2\si\in\mathbb{Z}$ and $\al-\beta\in\mathbb{Z}$ cannot be handled using these formula. Second, this transformation cannot be used for $u=\exp(\pm i\pi/3)=(1/2)(1\pm i\sqrt{3})$, because these points are mapped to themselves or each other, and at these points the serie diverges.\\
By considering  (\ref{y}) and (\ref{parameters}), the first problem means that $v$ has to be an integer and this is not the case for general physical parameters. In such an special cases, there is other transformation to tackle the problem \cite{nist}. But the second one is the major problem. Since the wave function is finite at these two points, the serie has to be terminated. The first possibility is  that $a$ or $b$ in the serie representation of $_2F_1$ in (\ref{F}) would be  equal to a negative integer. In the denominator of $A$ and $B$, if the functions $\Gamma(t)$'s  have a pole position at one of its arguments, then the exponentially growing term is absent.
 If $a=\al=-n$ then $B=0$, since it has the $\Gamma(\al)$ in its denominator. In other hand if we set $\beta=n'$ or $1-\beta=-n$, then $A=0$ for the same reason. Both possibilities have the same result as we will see below, i.e, if one of $A$ or $B$ vanishes, the other serie terminates.  Now, if we choose 
\be
1-\beta=-n, \quad n\in\mathbb{N}_0.
\ee\label{ev}
 then, $A=0$ and by using $\Gamma(z+n)=(z)_n\Gamma(z)$, $\Gamma(z-n)(z-n)_n=\Gamma(z)$ and $\Gamma(z)\Gamma(1-z)=\pi/\sin(\pi z)$, we get
\be\label{B}
B=(-1)^{n+1}\frac{\Gamma(-2\si+2)(-2\si+2)_n}{(4\si-2-2n)_n}.
\ee
From (\ref{ab}) and  the  condition $\beta=n+1$, we can find the discrete energy eigenvalues,
\be\label{ev1}
\epsilon_n=v-\left[\sqrt{\frac{1}{4}+v}-(n+\yd)\right]^2,
\ee
so that there is an upper bound for $n$,
\be\label{ub}
n+\yd\le\sqrt{v-\ep_n}.
\ee
(This condition is satisfied for finite values of $\la$ and for  the special case $\la\to\infty$,  $v\to 0$ and we have  
$\epsilon_n=-n^2$). Since for an atomic system $v$ is relative large number, the energy can be considered always positive.
If we use the physical parameters given in (\ref{def}), then we obtain
\be\label{ev2}
E_n=\frac{m\omega^2}{2\la}-\frac{\la\hb^2}{2m}
\left[\sqrt{\frac{1}{4}+\frac{m^2\omega^2}{\la^2\hb^2}}-(n+\yd)\right]^2.
\ee
And  in the limit $\la\to 0$, $E_n$ becomes
\be
\lim_{\la\to 0}E_n=(n+\yd)\hb\omega
\ee
as we have expected.
\subsection{Eigenfunctions of the Hamiltonian of NLHO}

Now we consider the equivalent representation of the solutions given in (\ref{sol2}) for the differential equation
(\ref{hg}) with  parameters given in (\ref{parameters}) and $c_2=0$,
\be
\varphi(z) =c_1u^{\si-1/4}(1-u)^{\si-1/4)}u^{\beta-\gamma}(u-1)^{\gamma-\al-\beta}\Big[ {_2F_1}(1-\beta,\gamma-\beta; \al-\beta+1;u^{-1})\Big].
\ee
By using the same transformation as given in (\ref{ac}), we get,
\begin{eqnarray*}
\!\!\!\!\!\!\!& &{_2F_1}(1-\beta,\gamma-\beta; \al-\beta+1;u^{-1})=\Gamma(\gamma-1)\Gamma(2-\gamma)
\Gamma(\al-\beta+1)\\
\!\!\!\!\! &\times&\!\!\!\! \Big[\frac{(-u^{-1})^{-(1-\beta)} }{\Gamma(\gamma-\beta)\Gamma(\al)}{ _2F_1}(1-\beta,1-\al; 2-\gamma; u)
- \frac{(-u^{-1})^{-(\gamma-\al)} }{\Gamma(1-\beta)\Gamma(\al-\gamma+1)}{ _2F_1}(\gamma-\beta,\gamma-\al; \gamma; u)\Big].
\end{eqnarray*}
The condition $1-\beta=-n$, implies $\Gamma(1-\beta)\to\infty$ and the second term in the bracket vanishes and the first hypergeometric function reduces to a polynomial of degree $n$. The coefficient of the first term  in RHS  is again $B$, as given in (\ref{AB}), so that we have the following eigenfunctions belonging to the eigenvalues $E_n$:
\be\label{EF}
\varphi_n(z)=-(-1)^{\al}c_1B u^{\si-1/4}(1-u)^{\si-1/4)}(1-u)^{\gamma-\al-\beta}u^{1-\gamma} { _2F_1}(1-\beta,1-\al; 2-\gamma; u).
\ee
The polynomial representation of ${_2F_1}$ is given by \cite{Abram},
\be
_2F_1(-n,a+1+b+n,a+1; u)=\frac{n!}{(a+1)_n}P_n^{(a,b)}(1-2u),
\ee
where $P_n^{(a,b)} $ is Jacobi's polynomial,
\be
P_n^{(a,b)}(1-2u)=\frac{\Gamma(a+n+1)}{n!\Gamma(a+b+n+1)}\sum_{m=0}^n(-1)^m\left(\begin{array}{c} n\\m\end{array}\right)\frac{\Gamma(a+b+n+m+1)}{\Gamma(a+m+1)}(u)^m.
\ee
By comparing to the relations  (\ref{EF}), (\ref{parameters}), (\ref{ux})  and the  condition $1-\beta=-n$, it is easy to see that: $1-2u=i\sinh z$,  $a=1-\gamma=1-2\si$, $b=\gamma-(\al+\beta)=1-2\si$,
and by using the relations (\ref{parameters}), (\ref{ev1}), we have $\al=4\si-n-2$.
Therefore the eigenfunctions (\ref{EF}) can be written as
\be\label{pn}
\varphi_n(z)=(-1)^{4\si-n-1}c_1Bu^{-\si+\frac{3}{4}}(1-u)^{-\si+\frac{3}{4}}
\frac{n!}{(-2\si+2)_n}P_n^{(-2\si+1,-2\si+1)}(-i\sinh z).
\ee
Thus in the old coordinates the eigenfunctions are given by
\be\label{pnx}
\varphi_n(x)=(-1)^{4\si-n-1}c_1B(1+\la x^2)^{-\si+\frac{3}{4}}
\frac{n!}{(-2\si+2)_n}P_n^{(-2\si+1,-2\si+1)}(-i\sla x).
\ee
and by Rodrigues' formula for special case $\sla x\in\mathbb{R}$
\[P_n^{(a,a)}(-i\sla x)=(\frac{-i}{\sla})^n\frac{(1+\la x^2)^{-a}}{2^nn!}\frac{d^n}{dx^n}\{(1+\la x^2)^{(a+n)}\}.\]
 In terms of physical parameters, $B$ is known and we could choose $c_1$ in such way that the eigenfunctions become normalized.
Now we calculate $\lim_{\la\to 0 }\varphi_n(x)$ in several steps:
\begin{itemize}
\item From (\ref{y}) we have
$\lim_{\la\to 0} \si\equiv\frac{1}{2\la b^2}=:\frac{\xi}{2}>0$, where $b^2:=\hb/m\omega$. So that
$\lim_{\la\to 0}(1-2\si)=-\xi$. From now on we use the $\lim_{\la\to 0} $ and $\lim_{\xi\to\infty}$ alternatively. 
\item We note that
$u^{-\si+\frac{3}{4}}(1-u)^{-\si+\frac{3}{4}}=(1/4)^{-\si+\frac{3}{4}}[1+\sinh^2\sqrt{\la}X]^{-\si+\frac{3}{4}}$. For large values of $\xi$, we have
$\sinh^2\sqrt{\la}X\approx \la x^2$, (thanks to  the canonical transformation between $X$ and $x$ as given in (\ref{trans})), so in this limit,
\[(1+\la x^2)^{-\xi/2}=
(1+\frac{1}{\xi/2}\frac{x^2}{2b^2})^{-\xi/2}=e^{-x^2/2b^2}.\]
\item There is a  relation between Jacobi and Hermite Polynomials \cite{Jose}
\[\lim_{\zeta\to\infty}\frac{2^nn!}{\zeta^{(n/2)}}P_n^{(\zeta,\zeta)}\left(y/\sqrt{\zeta}\right)=H_n(y),\]
where $\zeta\in\mathbb{R}$, and we have the same result for $\zeta\to-\infty$. Therefore if  we set
$\zeta=-\xi$, and considering that,
\[\lim_{\la\to 0} P_n^{-2\si+1,-2\si+1}(-i\sinh\sqrt{\la}X)= P_n^{-\xi,-\xi}\left(\frac{-ix}{-i\sqrt{-\xi b^2}}\right),\]
 then for large values of $\xi$, we have
 \[\lim_{\xi\to\infty}\frac{2^nn!}{(-\xi)^{(n/2)}}P_n^{-\xi,-\xi}(\frac{x}{\sqrt{-\xi b^2}})=H_n(x/b).\]
\end{itemize}
In the limit $\la\to 0$, by choosing  the appropriate form  for $c_1$,  we obtain the normalized  eigenfunctions  of the Hamiltonian of simple harmonic oscillator,
\be
\varphi_n(x)=\frac{1}{\sqrt{b}}\frac{1}{\sqrt{\pi^{1/2}2^n n!}}e^{-x^2/2b^2}H_n(\frac{x}{b}).
\ee

\section{Complexifier method}\label{4}
In this section we define complex coordinates on phase space and their corresponding operators by using Thiemann's method \cite{Thiem}.

\subsection{Classical complexifier}

The classical {\it complexifier}  is defined to be kinetic energy function divided by a constant of dimension frequence $\omega$, which can be expressed as
‚Äé‚Äé\be
\mbox{complexifier} =\frac{\mbox{kinetic}\;\; \mbox{energy}}{\omega}.
\ee
We construct complex-valued function $"z"$, corresponding to the position coordinate $x$, by taking the position functions $x$ and applying repeated Poisson brackets with the complexifier,
\be\label{a}
z=\sqrt{\frac{m\omega}{2}}\sum_{n=0}^\infty\frac{(i)^n}{n!}\{x,C\}_{(n)},
\ee
where Poisson bracket is given in (\ref{pois}) and  $\{x, C\}_{(0)}=x$, $\{x,C\}_{(n+1)}=\{\{x,C\}_{(n)},C\}$.
According to the original classical Hamiltonian (\ref{Hamil}), the kinetic energy is $(1+\la x^2)p^2/(2m)$ so that  the complexifier is defined by
\be
C=\frac{(1+\la x^2)p^2}{2m\omega}.
\ee
Some of the leading terms of Poisson bracket are given by,
 \[\{x,C\}_{(0)}=x,\quad \{x,C\}_{(1)}=\frac{1}{2m\omega}[2p(1+\la x^2)],\quad \{x,C\}_{(2)}=\frac{1}{(2m\omega)^2}[4p^2\la (1+\la x^2)x],\]
so that
\[\{x,C\}_{(2n)}=\frac{p^{2n}}{(m\omega)^{2n}}\la^{n}(1+\la x^2)^{n}x,\]
\[\{x,C\}_{(2n+1)}=\frac{p^{2n+1}}{(m\omega)^{2n+1}}\la^{n}(1+\la x^2)^{n+1}.\]
Then (\ref{a}) can be calculated,
\be
z=\sqrt{\frac{m\omega}{2}}\left[\cos\left(\frac{p\sqrt{\la(1+\la x^2)}}{m\omega}\right)x+\frac{i\sqrt{(1+\la x^2)}}{\sqrt{\la}}\sin\left(\frac{p\sqrt{\la(1+\la x^2)}}{m\omega}\right)\right].
\ee
In the limit $\la\to 0 $ we have
\be
\lim_{\la\to 0} z=\sqrt{\frac{m\omega}{2}}x+\frac{i}{\sqrt{2m\omega}}p.
\ee
The physical dimension of $z,z^\ast$ is $(angular\; momentum)^{(1/2)}$, so that their Poisson bracket is dimensionless.
They  are known as complex coordinates  on phase space and we also have
\begin{equation}
\lim_{\la\to 0}\omega z^\ast z=p^2/2m+m\omega^2 x^2/2.
\end{equation}
It is important to note that when $\la\not=0$,  this transformation is not canonical, i. e., in spite of  $\{x,p\}=1$,
\be\label{pz1}
\{z,z^\ast\}_{(x,p)}=-i\left[\la x^2+\cos\left(\frac{p\sqrt{\la(1+\la x^2)}}{m\omega}\right)\right],
\ee
and $\lim_{\la\to 0}\{z,z^\ast\}=-i$. \\
 We can find the corresponding complex coordinates for the canonical transformed $X, P$, according to  (\ref{trans}).
In the theory of  canonical transformation in classical mechanics we have the following statement:

 If $\phi:\mathbb{R}^{2f}\to\mathbb{R}^{2f}$ is a canonical transformation, then,
\[\{h\circ\phi,g\circ\phi\}=\{h,g\}\circ\phi,\]
where $f$ is degree of freedom, in our case $f=1$.
Accordingly,
\[\{x\circ\phi,C\circ\phi\}_{(n)}=\{x,C\}_{(n)}\circ\phi, \quad \phi(x,p)=(\frac{1}{\sqrt{\la}}\sinh\sqrt{\la}X,
\frac{P}{\cosh\sqrt{\la}X}).\]
Our new complex coordinates is given by
\ben\label{zz}
 Z&=&\sqrt{\frac{m\omega}{2}}\sum_{n=0}^\infty\frac{(i)^n}{n!}\{x,C\}_{(n)}\circ‚Äé\phi‚Äé\nonumber\\
 & =&\sqrt{\frac{m\omega}{2\la}}\left[\sinh(\sqrt{\la}X)\cos\left(\frac{\sqrt{\la}P}{m\omega}\right)+
 {i}\cosh(\sqrt{\la}X)\sin\left(\frac{\sqrt{\la}P}{m\omega}\right)\right].
\een
Therefore we get to the complex coordinate related to $x$,
\be\label{cz}
Z=\sqrt{\frac{m\omega}{2\la}}\sinh\left(\sqrt{\la}X+\frac{i\sqrt{\la}P}{m\omega}\right),‚Äé\quad‚Äé
Z^*=\sqrt{\frac{m\omega}{2\la}}\sinh\left(\sqrt{\la}X-\frac{i\sqrt{\la}P}{m\omega}\right).
\ee
In a similar way we get to the complex coordinate related to $X$, by using the same complexifier,
 \[A=\sqrt{\frac{m\omega}{2}}\sum_{n=0}^\infty\frac{(i)^n}{n!}\{X,C\}_{(n)}\]
and we also have
\be
A=\sqrt{\frac{m\omega}{2}}X+\frac{iP}{\sqrt{2m\omega}},\quad
 A^\ast=\sqrt{\frac{m\omega}{2}}X-\frac{iP}{\sqrt{2m\omega}}.
 \ee
 It is seen that $Z$ and $A$ have a similar relation, as $x$ and $X$,
 \be\label{za}
 Z=\sqrt{\frac{m\omega}{2\la}}\sinh\left(\sqrt{\frac{2\la}{m\omega}}A\right), \quad
 A=\sqrt{\frac{m\omega}{2\la}}\sinh^{-1}\left(\sqrt{\frac{2\la}{m\omega}}Z\right)
  \ee
  and there is a similar relation for $A^\ast $ and $Z^\ast$.

  {\bf Remark}. In comparison, these relations are counterparts of the coordinates $x$ and $X$ as given in  (\ref{trans}). Therefore  $Z$ is complexification of $x$, such as  $A$ is complexification  of   $X$,  so that  one can say, the complex coordinates are related according to the following diagram,
 \[ \begin{array}{ll}
 x\;\;\underrightarrow{Complexifier} & Z\\
\!\!\!\!\!\!\!ct\;\downarrow & \downarrow \\
  X\;\;\underrightarrow{Complexifier}&A
  \end{array}\]
where ct stand for  canonical transformation.
 By referring to (\ref{P}) we can see,
 \ben\label{da}
 \dot{A} &=& \{A, H\}\nonumber\\
 \dot{Z}   &=& \sqrt{\frac{\mo}{2}}\left(\frac{P}{m}-\frac{i\omega}{\sla}\frac{\sinh\sla X}{\cosh^3\sla X}\right)
 \cosh\left(\sqrt{\la}X+\frac{i\sqrt{\la}P}{m\omega}\right)\nonumber\\
  &=& \{Z,H\}.
 \een
  The Poisson bracket of $Z, Z^\ast$ leads to the same result as (\ref{pz1}),
\ben\label{pz2}
\{Z,Z^\ast\}_{(X,P)}&=&\frac{-i}{2}\left[\cosh(2\sqrt{\la X})+\cosh(\frac{2i\sqrt{\la}P}{\mo}\right]=-i\left[\la x^2+
\cos^2\left(\frac{p\sqrt{\la(1+\la x^2)}}{m\omega}\right)\right]\nonumber\\
&=& -i\cosh\left(\sqrt{\frac{2\la}{m\omega}}A\right)\cosh\left(\sqrt{\frac{2\la}{m\omega}}A^\ast\right)\nonumber\\
&=& \{Z,Z^\ast\}_{(x,p)}=-i\{Z,Z^\ast\}_{(A,A^\ast)}\nonumber\\
\{A,A^\ast\}_{(X,P)}&=&-i=\{A,A^\ast\}_{(x,p)}.
\een

 \subsection{Quantum Mechanical Complexifier}

Analogously, the complexifier in quantum mechanics is defined by kinetic energy operator, so that the corresponding operator of complex coordinate is given by
\be \label{qz}
\hat{z}=\sqrt{\frac{m\omega}{2}}\sum_{n=0}^\infty\frac{(i)^n}{n!}\frac{[\hat{x},\hat{C}]_{(n)}}{(i\hb)^n}=
e^{-\hat{C}/\hb}\hat{x}e^{\hat{C}/\hb},
\ee
where $\hat{z}$ plays the role of {\it annihilation operator} and as given in  (\ref{ke}),
\be
\hat{C}=\frac{-\hb^2}{2m\omega}\left[\yd(1+\la x^2)\frac{\p^2}{\p x^2}+\yd\la x\frac{\p}{\p x}\right],
\ee
and $[\hat{x}, \hat{C}]_{(0)}=\hat{x}$, $[\hat{x},\hat{C}]_{(n+1)}=[[\hat{x},\hat{C}]_{(n)},\hat{C}]$.
The calculation of different order of commutator in $(x,p)$-coordinate system is cumbersome. It is easier to go to the canonical transformed coordinates $X, P$. But again we have to be careful in definition of kinetic energy operator.

As it is done in the case of classical complexifier, the equivalent version of (\ref{qz}) is given by
\ben\label{QZ}
\hat{Z}&=&\sqrt{\frac{m\omega}{2}}\sum_{n=0}^\infty \frac{(i)^n}{(i\hb)^nn!}
\left[\frac{1}{\sqrt{\la}}\sinh\sqrt{\la}X\;,\;\frac{-\hb^2}{2m\omega}\frac{d^2}{dX^2}\right]_{(n)}\nonumber\\
&=&\sqrt{\frac{m\omega}{2\la}}\sum_{n=0}^\infty \left(\frac{\eta}{2}\right)^n\frac{(-1)^n}{n!}
\left[\sinh\sqrt{\la}X\;,\;\frac{d^2}{dX^2}\right]_{(n)},
\een
where $\eta=\frac{\hb}{m\omega}.$ The following commutation relations will be useful in sequel,
\[ \left[ \sinh\sqrt{\la}X\;, \frac{d^2}{dX^2}\right]=-\la \sinh\sqrt{\la}X-2\sqrt{\la}\cosh\sqrt{\la}X\frac{d}{dX}, \]
\[ \left[ \cosh\sqrt{\la}X,\frac{d^2}{dX^2}\right]=-\la \cosh\sqrt{\la}X-2\sqrt{\la}\sinh\sqrt{\la}X\frac{d}{dX}. \]

Therefore we can calculate the repeated commutators more easier. Some of the leading terms in summation read as

{\tiny \[ S_0=\sinh\sqrt{\la}X\]
\[S_1=(\frac{\eta}{2})\left((\sinh\sqrt{\la}X)\la+(\cosh\sqrt{\la}X)2\sqrt{\la}\frac{d}{dX}\right)\]
\[S_2=(\frac{\eta}{2})^2\frac{1}{2!}\left((\sinh\sqrt{\la}X)(\la^2+4\la\frac{d^2}{dX^2})+
(\cosh\sqrt{\la}X)(4\la^{3/2}\frac{d}{dX})\right)\]
\[S_3=(\frac{\eta}{2})^3\frac{1}{3!}\left((\sinh\sqrt{\la}X)(\la^3+12\la^2\frac{d^2}{dX^2})+
(\cosh\sqrt{\la}X)(6\la^{5/2}\frac{d}{dX}+8\la^{3/2}\frac{d^3}{dX^3})\right)\]
\[S_4=(\frac{\eta}{2})^4\frac{1}{4!}\left((\sinh\sqrt{\la}X)(\la^4+24\la^3\frac{d^2}{dX^2}+16\la^2\frac{d^4}{dX^4})+
(\cosh\sqrt{\la}X)(8\la^{7/2}\frac{d}{dX}+32\la^{5/2}\frac{d^3}{dX^3})\right)\]
{\tiny\[S_5=(\frac{\eta}{2})^5\frac{1}{5!}\left((\sinh\sqrt{\la}X)(\la^5+40\la^4\frac{d^2}{dX^2}+80\la^3\frac{d^4}{dX^4})+
(\cosh\sqrt{\la}X)(10\la^{9/2}\frac{d}{dX}+80\la^{7/2}\frac{d^3}{dX^3}+32\la^{5/2}\frac{d^5}{dX^5} )\right)\]}
{\tiny\[S_6=(\frac{\eta}{2})^6\frac{1}{6!}\left((\sinh\sqrt{\la}X)(\la^6+60\la^5\frac{d^2}{dX^2}+240\la^4\frac{d^4}{dX^4}
+64\la^3\frac{d^6}{dX^6})+
(\cosh\sqrt{\la}X)(12\la^{11/2}\frac{d}{dX}+160\la^{9/2}\frac{d^3}{dX^3}+192\la^{7/2}\frac{d^5}{dX^5} )\right)\]}
{\tiny\[S_7=(\frac{\eta}{2})^7\frac{1}{7!}\left((\sinh\sqrt{\la}X)(\la^7+84\la^6\frac{d^2}{dX^2}+560\la^5\frac{d^4}{dX^4}
+448\la^4\frac{d^6}{dX^6})+
(\cosh\sqrt{\la}X)(14\la^{13/2}\frac{d}{dX}+280\la^{11/2}\frac{d^3}{dX^3}+672\la^{9/2}\frac{d^5}{dX^5}+
128\la^{7/2}\frac{d^7}{dX^7} )\right).\]}}
Then after summation over appropriate terms we get,
\ben\label{z1}
\hat{Z}&=&\sqrt{\frac{m\omega}{2\la}}e^{\eta\la/2}\left[\sinh(\sqrt{\la} X)\cosh(\eta\sqrt{\la}\frac{d}{dX})+
\cosh(\sqrt{\la} X)\sinh(\eta\sqrt{\la}\frac{d}{dX})\right]\nonumber\\
&=&\sqrt{\frac{m\omega}{2\la}}e^{\hb\la/2m\omega}\left[\sinh(\sqrt{\la} X)\cos(\frac{\sqrt{\la}\hat{P}}{m\omega})+
i\cosh(\sqrt{\la} X)\sin(\frac{\sqrt{\la}\hat{P}}{m\omega})\right].
\een
Here we have to note that again, like the classical complexifier, $\hat{Z}$ is complex operator corresponding to the  $\hat{x}$. If we construct complexification of $\hat{X}$ by using  $\hat{C}$ we get,
\ben\label{ca}
\hat{A} &=&\sqrt{\frac{m\omega}{2\hb}}\sum_{n=0}^\infty \frac{(i)^n}{(i\hb)^nn!}
\left[\hat{X}\;,\;\frac{-\hb^2}{2m\omega}\frac{d^2}{dX^2}\right]_{(n)}\nonumber\\
&=&\sqrt{\frac{m\omega}{2\hb}}\hat{X}+\frac{i\hat{P}}{\sqrt{2m\omega\hb}},
\een
where we have multiplied the factor $1/\sqrt{\hb}$,  to make $\hat{A}$  a dimensionless operator.
 By using the Bakerñ Campbellñ Hausdorff (BCH) formula 
 \[e^{(A+B)}=e^Ae^Be^{-[A,B]/2},\; \mbox{if} \quad [A,[A,B]]=0=[B,[A,B]]\]
 and writing the hyperbolic and trigonometric  functions in  exponential representation,
 \be
  \hat{Z}=\sqrt{\frac{m\omega}{2\la}}e^{\la \hb/m\omega}\left[\sinh(\sqrt{\la}\hat{X}+\frac{i\sqrt{\la}\hat{P}}{m\omega})\right], \quad
  \hat{Z}^\dagger=\sqrt{\frac{m\omega}{2\la}}e^{\la \hb/m\omega}\left[\sinh(\sqrt{\la}\hat{X}-\frac{i\sqrt{\la}\hat{P}}{m\omega})\right]   \ee
 we can see the correspondence between classical, (\ref{cz}), and quantum complex coordinates. Although in the quantum case, by considering  the  operators ordering problem, we were   careful in using the trigonometric combination formula  but thanks to the properties of these functions, we see this correspondence, up to a constant factor, nicely. 
 The result commutator, then, will be achieved,
\be
[\hat{Z}, \hat{Z}^\dagger]=\frac{m\omega}{2\la}e^{2\la \hb/m\omega}\sinh(\frac{\hb\la}{m\omega})\left[\cosh(2\sla \hat{X})+\cosh(\frac{2\sla\hat{iP}}{m\omega})\right].
\ee
Again, in the limit $\la\to 0$,
\[{\lim_{\la\to 0}}[\hat{Z}, \hat{Z}^\dagger]=\hb,\]
and by referring to the equations (\ref{pz1}, \ref{pz2})
\[\lim_{\la\to 0}\frac{1}{i\hb}[\hat{Z}, \hat{Z}^\dagger]=\{Z,Z^\ast\}=\{z,z^\ast\},\]
and
\be \yd(\hat{Z} \hat{Z}^\dagger+ \hat{Z}^\dagger\hat{Z})=\frac{m\omega}{2\la}e^{2\la b^2}\cosh \la b^2
\left(\cosh^2\sla X-\cos^2(\sla b^2P/\hb )\right).
\ee
According to the Hamiltonian (\ref{SE}) it easy to see that
\be
\dot{\hat{A}}=\frac{1}{i\hb}[\hat{A}, \hat{H}],\quad  \dot{\hat{Z}}=\frac{1}{i\hb}[\hat{Z}, \hat{H}].
\ee
Again we see the correspondence between quantum  and classical, (\ref{da}), equations  of motions.
\section{Three types of coherent states}\label{5}

\subsection{Type 1: Complexifier coherent states}
At first,  one can immediately recognize that by using  (\ref{ca}),
\be
\frac{\hb\omega}{2}(\hat{A}^\dagger \hat{A}+\hat{A}\hat{A}^\dagger)=\frac{\hat{P}^2}{2m}+\yd m\omega^2\hat{X}^2,
\quad \text{and}~~~[ \hat{A}, \hat{A}^\dagger]=1.
\ee
It is easy to show that
\be
 \hat{A}\psi_0(X')=0,\quad \psi_0(X')=\left( \frac{1}{\pi b^2}\right)^{\yc}\exp(-\frac{X'^2}{2b^2}),
\ee
in the limit $\la\to 0$ we get the ground state of SHO. In the old coordinates
\be
\psi_0(x')=\left( \frac{1}{\pi b^2}\right)^{\yc}\exp\left(-\frac{(\sinh^{-1}\sla x')^2}{2\la b^2}\right),
\ee
where $b^2=\hb/(m\omega)$, and for every $\ga\in\mathbb{C}$, we can easily construct other eigenfunctions of
$\hat{A}$, by using the displacement operator,
\[ \lef X'|\ga\ra=e^{-|\ga|^2/2}e^{\ga \hat{A}^\dagger}\psi_0(X').\]
Now, we have the standard coherent states in $(X,P)$-coordinates,
\[ \psi_\ga(X')=\lef X'|\ga\ra=e^{\frac{-i}{2\hb}\lef\hat{P}\ra_\ga\lef\hat{X}\ra_\ga}\left(\frac{m\omega}{\pi\hb}\right)^\yc\exp\left\{-\frac{(X'-\lef\hat{X}\ra_\ga)^2}{2\hb/(m\omega)}\right\}
e^{\frac{i}{\hb}\lef\hat{P}\ra_\ga X'}\]
\[\lef \hat{X}\ra_\ga=\sqrt{2}\,b\,\Re \ga,\quad \lef \hat{P}\ra_\ga=\sqrt{2}\,bm\omega\,\Im \ga,\]
and it can be shown $\hat{A}|\ga\ra=\ga |\ga\ra$. We have also  the following relations, 
\[\frac{\hat{Z}}{\sqrt{\hb}}=:\hat{Z}'=\sqrt{\frac{m\omega}{2\la\hb}}e^{\la b^2}\sinh[b\sqrt{2\la}\hat{A}],
\quad \hat{A}=\frac{1}{b\sqrt{2\la}}\sinh^{-1}[e^{-\la b^2}\sqrt{(2\la/\hb)}b\hat{Z}].\]
 which are similar to (\ref{za}). We have divided both sides by $\sqrt{\hb}$, for dimensional convention.   Then 
\[\hat{Z'}|\ga\ra=\sqrt{\frac{1}{2\la b^2}}e^{\la b^2}\sinh[b\sqrt{2\la}\ga]|\ga\ra\equiv f(\ga)|\ga\ra\]

To find the explicit form of the eigenfunctions of $\hat{Z}$, we assume that $|z\ra$ is the desired eigenket, 
$\hat{Z'}|z\ra=z|z\ra$. Using the (over)completeness of $|\ga\ra$'s, $\pi^{-1}\int d^2\ga |\ga\ra\lef\ga|=1$, and the inner product of standard coherent states $|\lef z|\ga\ra|^2=\exp(-|z-\ga|^2)$,  
$z$ is given by the following equation, 
\[z=\pi^{-1}\int d^2\ga f(\ga)|\lef z|\ga\ra|^2=\frac{e^{\la b^2}}{\sqrt{2\la}\pi b}\int d^2\ga \sinh[b\sqrt{2\la}\ga]
e^{-|z-\ga|^2}\]
\subsection{Type 2: f-deformed  coherent states}

Now we introduce another form for energy eigenvalues (\ref{ev2}). We have to find a  $n$-dependent function $f(n)$, so that $E_n$ is written as,
\be
E_n=\frac{\hb\omega}{2}\left[(n+1)|f(n+1)|^2+n|f(n)|^2\right].
\ee
This function is given by
\be
f(n)=\sqrt{\frac{\la\hb}{m\omega}}\sqrt{(\frac{1}{4}+
\frac{m^2\omega^2}{\la^2\hb^2})^{(1/2)}-\frac{n}{2}}=\left(\sqrt{(\frac{1}{4v}+1)}-\frac{n}{2\sqrt{v}}\right)^{1/2}.
\ee
According to the upper bound condition (\ref{ub}), we can consider the $f(n)$ as a real function;
and we have
\be
 \lim_{\la\to 0}f(n)=1, \qquad \lim_{f(n)\to 1}E_n=(n+\yd)\hb\omega.
 \ee
First we give the following representation for the Hamiltonian in Fock space of  SHO, by using the relation
(\ref{ev2}),

\ben
\hat{H} &=&\yd(\hat{b}\hat{b}^\dagger+\hat{b}^\dagger\hat{b})\nonumber\\
&=&\frac{\hb\omega}{2}\left[(\hat{n}+1)|f(\hat{n}+1)|^2+\hat{n}|f(\hat{n})|^2\right]\nonumber\\
&=& \frac{\hb\omega}{2\sqrt{v}}\left(v-\left[\sqrt{\frac{1}{4}+v}-(\hat{n}+\yd) \right]^2\right).
\een
Then, we can formally construct the  relations between creation and annihilation operators of NLHO and SHO,
\be
\hat{b}=\hat{a}f(\hat{n}),\quad \hat{b}^\dagger=f(\hat{n})\hat{a}^\dagger,\quad \hat{a}^\dagger \hat{a}=\hat{n},
\ee
so that
\ben
[\hat{b},\hat{b}^\dagger] &=& \frac{\hb\omega}{2}\left[(\hat{n}+1)|f(\hat{n}+1)|^2-\hat{n}|f(\hat{n})|^2\right]\nonumber\\
&=& \frac{\hb\omega}{2\sqrt{v}}\left[ \sqrt{\frac{1}{4}+v}- (\hat{n}+\yd)\right].
\een
The coherent states are given as right hand eigenfunctions of the  new annihilation operator
\be
\hat{b}|f,\beta\ra=\beta |f,\beta\ra.
\ee
\subsection{Type 3: Generalized displacement operator}

The  Hamiltonian (\ref{SE}) can be easily factorized in the following way.
\be\label{BB}
\hat{H}=\yd(\hat{B}^\dagger\hat{B}+\hat{B}\hat{B}^\dagger),
\ee
where
\be
\hat{B}=\left[\frac{\hb}{\sqrt{2m}}\frac{d}{dX}+\sqrt{\frac{m\omega^2}{2\la}}\tanh(\sla X)\right], \quad
\hat{B}^\dagger=\left[\frac{-\hb}{\sqrt{2m}}\frac{d}{dX}+\sqrt{\frac{m\omega^2}{2\la}}\tanh(\sla X)\right],
\ee
and
\be
\frac{1}{\hb\omega}[\hat{B},\hat{B}^\dagger]=\frac{1}{\cosh^2(\sla X)}=\frac{1}{1+\la x^2}.
\ee
The ground state equation for annihilation operator $\hat{B}\psi_0(X)=0$, leads to
\be
\psi_0=c_1[\cosh(\sla X)]^{(-1/\la b^2)}=c_1[\sqrt{1+\la x^2}]^{(-1/\la b^2)}.
\ee
If we define $\xi=1/(2\la b^2)$, so that for $\la\to 0$,  $\xi\to \infty$, then we have
\[\lim_{\la\to  0}\psi_0(x)=\lim_{\xi\to \infty} c_1(1+\frac{1}{\xi}\frac{x^2}{2b^2})^{-\xi}=c_1e^{-x^2/2b^2}.\]
This is the ground state of simple harmonic oscillator.
Since this system belongs to the shape invariant class, the ground state wave function should be of the form $\phi_0(X)=c_2[\cosh(\sla X)]^{(-\al E_0)}$, so that
\[\hat{H}\phi_0(X)=E_0\phi_0(X), \quad E_0=\frac{\la\hb^2}{2m}\left[ -\yd+\sqrt{\yc +\frac{1}{\la^2b^4}}\right],\quad\text{with}~~~
\al=\frac{2m}{\la\hb^2},\]
where $\phi_0(X)$ and $E_0$ are  exactly  the same as we have   in (\ref{pnx}) and  (\ref{ev2}).
Again, for $\la\to 0$, $E_0\to(\hb\omega/2)$, and $|\phi_0\ra\to |0\ra$, i.e., the ground state of SHO.
Now, we construct the generalized displacement operator,
\be
\hat{D}(\zeta)=e^{\zeta \hat{B}^\dagger-\zeta^\ast \hat{B}}.
\ee
And the coherent states are given by
\be
D(\zeta)|\phi_0\ra=|\zeta\ra, \quad \text{and}~~~\hat{B}|\zeta\ra=\zeta|\zeta\ra.
\ee
\section*{Conclusion and final remarks}
Three types of annihilation operators and their eigenfunctions, i.e.,   corresponding  coherent states, are introduced for 1D NLHO. All of them are dependent  on a deformation parameter $\la$, which can be realized as a control parameter in  different physical problem  (as we will see in the second part of this work), and  for $\la\to 0$, they are identical with the  annihilation operator of SHO.

The most important problems, we are going to address in the second part of this work, are: Presentation of  the explicit form of coherent states and investigation  their most important properties, such as resolution of the identity, continuity in label (since the coherent states will be labelled by points of the  phase space), and  the relationship between these coherent states, by exploring the properties  of Jacobi polynomials $P_n^{\al,\beta}$ as eigenfunctions of the Hamiltonian of NLHO. They play the role of  Hermitian polynomials $H_n$ in the theory of SHO, so that the corresponding coherent states may be constructed by their superposition, in a similar way as the  standard coherent states of SHO are constructed out of $H_n$'s.

\section*{Acknowledgment}
R. R. wishes to thank Vice President of Research and Tecknology  of University of Isfahan, for financial support  and Albanova University Center of Stockholm for hospitality. \\ H. H. thanks  the Swedish Research Council (VR) for support.

\end{document}